\documentclass[12pt]{article}
\usepackage{amssymb}
\usepackage{amsmath}
\usepackage[dvips]{graphicx}
\usepackage{fancyhdr}
\usepackage{xcolor}
\usepackage{ifpdf}
\usepackage{graphicx}

\usepackage{color}

\allowdisplaybreaks  

\title{Holographic Q-picture of Kerr-Newman-AdS-dS Black Hole}

\author{Bin Chen$^{1,2}${\footnote{Email:bchen01@pku.edu.cn}},
Chiang-Mei Chen$^3$\footnote{Email:cmchen@phy.ncu.edu.tw}, and
Bo Ning$^1${\footnote{Email:ningbo@pku.edu.cn}}
\\
{\small $^{1}$Department of Physics, and State Key Laboratory of Nuclear Physics and Technology,} \\
{\small Peking University, Beijing 100871, P.R. China}
\\
{\small $^{2}$Center for High Energy Physics, Peking University, Beijing 100871, P.R. China}
\\
{\small $^3$Department of Physics and Center for Mathematics and Theoretical Physics,} \\
{\small National Central University, Chungli 320, Taiwan}}

\begin{document}

\maketitle

\begin{abstract}

In this article, we show that a four-dimensional Kerr-Newman-AdS-dS black hole could be described by two different holographic two-dimensional conformal field theories. The first description, in view of the angular momentum, has been well discussed in~\cite{Chen:2010bh}. The other alternative dual picture, however, is somehow overlooked. This new description, in view of the charge, provides another holographic correspondence. This picture again is supported not only by the match of the macroscopic entropy with the counting of microscopic degrees of freedom via the Cardy formula and also by the agreement on the real-time correlator in the superradiant scattering.

\end{abstract}
\newpage

\section{Introduction}

It is a remarkable fact that an asymptotically flat four dimensional Kerr black hole could be described holographically by a two-dimensional conformal field theory (CFT). The Kerr/CFT correspondence was first proposed for the extremal Kerr black hole by studying the asymptotic symmetry group of its near horizon geometry~\cite{AndyWei, matsuo, Castro:2009jf}. This correspondence was firstly supported by the match of the microscopic entropy counting with the macroscopic Bekenstein-Hawking entropy, and moreover supported by the agreement of the superradiant scattering amplitudes with the dual CFT prediction~\cite{Bredberg:2009pv, Hartman:2009nz, Cvetic:2009jn, ChenChu, Becker:2010jj}. Furthermore, the Kerr/CFT correspondence was generalized to a generic non-extremal Kerr black hole by the study of the hidden conformal symmetry acting on the solution space~\cite{Castro:2010fd}. It was found that in the low-frequency limit the radial equation of the scalar scattering off a Kerr black hole could be written in terms of $SL(2, R)$ quadratic Casimir. Though the hidden conformal symmetry can not generate new solutions, it really determines the scattering amplitudes. Both the entropy counting and the low frequency amplitudes again support the holographic picture.

Inspired by the Kerr/CFT correspondence, the holographic descriptions of other kinds of black holes have been investigated~\cite{KerrCFT, HiddenSymmetry}. Among them, the holographic description of the Reissner-Nordstr\"om~(RN) charged black hole is of particular interest. For the extremal RN black hole, the charge gives the radius of the AdS space appearing in the near horizon geometry, and therefore is expected to determine the central charge of the dual CFT~\cite{Hartman:2008pb, Garousi:2009zx, Chen:2009ht, Chen:2010bsa, Chen:2010yu, Chen:2010as}. However, unlike the Kerr case, the AdS$_3$ information in RN black hole is not completely encoded in the geometry, but also in the gauge potential, therefore it is subtle to compute the central charge of the dual CFT~\cite{Chen:2010bsa}. The Kerr/CFT and RN/CFT correspondences naturally suggest that for the four-dimensional Kerr-Newman black hole there should exist two distinct hidden conformal symmetries associated with twofold holographic descriptions, called J-picture and Q-picture respectively~\cite{Chen:2010yw}. This is reminiscent of the multi-fold descriptions of higher-dimensional extremal Kerr black holes which have multiple independent angular momenta~\cite{Lu:2008jk, Chen:2009xja, Krishnan:2010pv}.

On the other hand, the investigation of holographic pictures of the Kerr black holes in AdS or dS spacetime is also remarkable. In this case, the function determining the horizon is quartic rather than quadratic for asymptotically flat black hole and this fact causes the search for the hidden conformal symmetry be more tricky. Different from the Kerr case in which one can move away from the horizon and work with the ``near'' region, one has to focus strictly on the near horizon region to find the hidden conformal symmetry. This is in accordance with the universal behavior of the black hole. In~\cite{Chen:2010bh}, a holographic picture, in view point of angular momentum, of the Kerr(-Newman)-AdS-dS black hole has been proposed. Both the Bekenstein-Hawking entropy and the superradiant scattering amplitudes are in good agreement with the CFT prediction.

It is a natural expectation that there should be a Q-picture description dual to a Kerr-Newman-AdS-dS black hole. In this article, we will explore this picture in more details. In the next section, we discuss the hidden conformal symmetry in holographic Q-picture. Similar to the treatment in~\cite{Chen:2010bh}, we need to focus on the near-horizon region. After turning off the angular mode of the probe scalar field, we find that the radial equation could be rewritten in terms of $SL(2, R)$ quadratic Casimir, indicating the existence of a hidden conformal symmetry. As a result, we read out the temperatures of dual CFT. In section 3, we present the microscopic description in Q-picture. The key theme is to find the central charges of the dual CFT. As in the other cases, we first consider the central charges of the near-extremal black holes and expect the same expression holds even for generic black holes. In the Q-picture, another subtle point is that we have to uplift the 4D black hole to 5D in order to read the central charges.  It turns out that the central charges and temperatures include a free parameter. Furthermore we discuss the thermodynamics of the black hole and calculate the superradiant scattering amplitude, which is in perfect agreement with the CFT prediction. We end with some discussions in section 4.

\section{Hidden conformal symmetry in Q-picture}

For a four-dimensional Kerr-Newman-AdS-dS black hole, its metric takes the following form in Boyer-Lindquist-type coordinates~\cite{Caldarelli:1999xj}
\begin{equation} \label{KerrNewman}
ds^2 = - \frac{\Delta_r}{\rho^2} \left( d t - \frac{a \sin^2\theta}{\Xi} d\phi \right)^2 + \frac{\rho^2}{\Delta_r} dr^2 + \frac{\rho^2}{\Delta_\theta} d\theta^2 + \frac{\Delta_\theta}{\rho^2} \sin^2\theta \left( a dt - \frac{r^2 + a^2}{\Xi} d\phi \right)^2,
\end{equation}
where
\begin{eqnarray}
\Delta_r &=& (r^2 + a^2) \left( 1 + \frac{r^2}{l^2} \right) - 2 M r + q^2, \qquad q^2 = q^2_e + q^2_m,
\nonumber\\
\Delta_\theta &=& 1 - \frac{a^2}{l^2} \cos^2\theta,
\nonumber\\
\rho^2 &=& r^2 + a^2 \cos^2\theta,
\nonumber\\
\Xi &=& 1 - \frac{a^2}{l^2}.
\end{eqnarray}
Here $l^{-2}$ is the normalized cosmological constant, which is positive for dS and negative for AdS. The above metric reduces to the one of a Kerr-Newman black hole when $l^{-2} = 0$, and it describes a RN-AdS-dS black hole when $a = 0$. The physical mass, angular momentum and charges of the black hole are related to the parameters $M, a, q_{e, m}$ by
\begin{equation}
M_{ADM} = \frac{M}{\Xi^2}, \qquad J = \frac{a M}{\Xi^2}, \qquad Q_{e, m} = \frac{q_{e, m}}{\Xi}.
\end{equation}
The gauge potential and its field strength are respectively
\begin{eqnarray}
A_{[1]} &=& - \frac{q_e r}{\rho^2} \left( d t - \frac{a \sin^2\theta}{\Xi} d\phi \right) - \frac{q_m \cos\theta}{\rho^2} \left( a dt - \frac{r^2 + a^2}{\Xi} d\phi \right),
\\
F_{[2]} &=& - \frac{q_e (r^2 - a^2 \cos^2\theta) + 2 q_m r a \cos\theta}{\rho^4} \left( d t - \frac{a \sin^2\theta}{\Xi} d\phi \right) \wedge dr
\nonumber\\
& & + \frac{q_m (r^2 - a^2 \cos^2\theta) - 2 q_e r a \cos\theta}{\rho^4} \sin\theta d\theta \wedge \left( a dt - \frac{r^2 + a^2}{\Xi} d\phi \right).
\end{eqnarray}

In the following, for simplicity, we just focus on the electrically charged black hole, i.e. $q_m = 0, \; q = q_e$. The Hawking temperature, entropy and angular velocity of the horizon are respectively
\begin{eqnarray}
T_H &=& \frac{r_+}{4 \pi (r_+^2 + a^2)} \left( 1 - \frac{a^2 + q^2}{r^2_+} + \frac{a^2}{l^2} + \frac{3 r^2_+}{l^2} \right),
\nonumber\\
S_{BH} &=& \frac{\pi (r_+^2 + a^2)}{\Xi},
\nonumber\\
\Omega_H &=& \frac{a \Xi}{r_+^2 + a^2}. \label{OH}
\end{eqnarray}
The electric potential $\Phi$, measured at infinity with respect to the horizon, is
\begin{equation}
\Phi_e = A_\mu \xi^\mu \biggr|_{r \to \infty} - A_\mu \xi^\mu \biggr|_{r=r_+} = \frac{q_e r_+}{r^2_+ + a^2}, \label{Epotential}
\end{equation}
where $\xi = \partial_t + \Omega_H \partial_\phi$ is the null generator of the horizon.

As discussed in~\cite{Chen:2010bh}, the scattering issue in the Kerr-AdS-dS and Kerr-Newman-AdS-dS black holes needs a careful treatment, because that the function $\Delta_r$ is quartic. In order to find the the hidden conformal symmetry, one has to focus on the near-horizon region. In the near-horizon region, the function $\Delta_r$ can be expanded to the quadratic order of $r - r_+$,
\begin{eqnarray}
\Delta_r &=& (r^2 + a^2) \left( 1 + \frac{r^2}{l^2} \right) - 2 M r + q^2
\nonumber\\
&\simeq & k (r - r_+) (r - r_\ast),
\end{eqnarray}
where $r_+$ is the outer horizon, and
\begin{eqnarray}
k &=& 1 + \frac{a^2}{l^2} + \frac{6 r^2_+}{l^2}, \label{k}
\\
r_\ast &=& r_+ - \frac{r_+}{k} \left( 1 - \frac{a^2 + q^2}{r^2_+} + \frac{a^2}{l^2} + \frac{3 r^2_+}{l^2} \right). \label{rstar}
\end{eqnarray}

For a complex massless scalar field with charge $e$, its dynamics is given by the Klein-Gordon (KG) equation
\begin{equation}
(\nabla_\mu + i e A_\mu) (\nabla^\mu + i e A^\mu) \Phi = 0.
\end{equation}
By imposing the following ansatz
\begin{equation} \label{ansatz}
\Phi = e^{-i \omega t + i m \phi} {\cal S}(\theta) {\cal R}(r),
\end{equation}
the KG equation is decoupled into an angular equation
\begin{equation}
\frac{1}{\sin\theta} \partial_\theta \left( \sin\theta \Delta_\theta \partial_\theta {\cal S} \right) - \frac{(m \Xi)^2}{\Delta_\theta \sin^2\theta} {\cal S} + \frac{2 m a \Xi \omega - a^2 \omega^2 \sin^2\theta}{\Delta_\theta} {\cal S} + K {\cal S} = 0,
\end{equation}
and a radial equation
\begin{equation}
\partial_r \left( \Delta_r \partial_r {\cal R} \right) + \frac{[ \omega (r^2 + a^2) - m a \Xi - e q_e r ]^2}{\Delta_r} {\cal R} - K {\cal R} = 0,
\end{equation}
where $K$ is the separation constant. In the near-horizon region and in the low frequency limit $r_+ \omega \ll 1$, the radial equation could be simplified as
\begin{equation} \label{scalarKNAdS}
\partial_r [ (r - r_+) (r - r_\ast) \partial_r {\cal R} ] + \frac{r_+ - r_\ast}{r - r_+} A \, {\cal R} + \frac{r_+ - r_\ast}{r - r_\ast} B \, {\cal R} + C \, {\cal R} = 0,
\end{equation}
with
\begin{eqnarray}
A &=& \frac{\left[ \omega (r^2_+ + a^2) - m a \Xi - e q_e r_+ \right]^2}{k^2 (r_+ - r_\ast)^2},
\nonumber\\
B &=& - \frac{\left[ \omega (r^2_\ast + a^2) - m a \Xi - e q_e r_\ast \right]^2}{k^2 (r_+ - r_\ast)^2},
\nonumber\\
C &=& \frac{e^2 q_e^2}{k^2} - \frac{K}{k}.
\end{eqnarray}

In order to study the hidden conformal symmetry, we need to introduce the following conformal coordinates for non-extremal black holes:
\begin{eqnarray}
\omega^+ &=& \sqrt{\frac{r - r_+}{r - r_\ast}} \; e^{2 \pi T_R \phi + 2 n_R t},
\nonumber\\
\omega^- &=& \sqrt{\frac{r - r_+}{r - r_\ast}} \; e^{2 \pi T_L \phi + 2 n_L t},
\nonumber\\
y &=& \sqrt{\frac{r_+ - r_\ast}{r - r_\ast}} \; e^{\pi (T_L + T_R) \phi + (n_L + n_R) t},
\nonumber
\end{eqnarray}
from which we may define locally the vector fields
\begin{eqnarray}
H_1 &=& i \partial_+,
\nonumber\\
H_0 &=& i \left( \omega^+ \partial_+ + \frac{1}{2} y \partial_y \right),
\nonumber\\
H_{-1} &=& i \left( \omega^{+2} \partial_+ + \omega^+ y \partial_y - y^2 \partial_- \right),
\end{eqnarray}
and
\begin{eqnarray}
\tilde H_1 &=& i \partial_-,
\nonumber\\
\tilde H_0 &=& i \left( \omega^- \partial_- + \frac{1}{2} y\partial_y \right),
\nonumber\\
\tilde H_{-1} &=& i \left( \omega^{-2} \partial_- + \omega^- y \partial_y - y^2 \partial_+ \right).
\end{eqnarray}
These vector fields obey the $SL(2,R)$ Lie algebra
\begin{equation}
[ H_0, H_{\pm 1} ] = \mp i H_{\pm 1}, \qquad [H_{-1}, H_1] = -2 i H_0,
\end{equation}
and similarly for $(\tilde H_0, \tilde H_{\pm 1})$. The quadratic Casimir is
\begin{eqnarray}
{\cal H}^2 = \tilde{\cal H}^2 &=& - H_0^2 + \frac{1}{2} (H_1 H_{-1} + H_{-1} H_1)
\nonumber\\
&=& \frac{1}{4} (y^2 \partial^2_y - y \partial_y) + y^2 \partial_+ \partial_-.
\end{eqnarray}
In terms of $(t, r, \phi)$ coordinates, the Casimir becomes
\begin{eqnarray} \label{H2}
{\cal H}^2 &=& (r - r_+) (r - r_\ast) \partial_r^2 + (2 r- r_+ - r_\ast) \partial_r
\nonumber\\
&& + \frac{r_+ - r_\ast}{r - r_\ast} \left( \frac{n_L - n_R}{4 \pi G} \partial_\phi - \frac{T_L - T_R}{4 G} \partial_t \right)^2
\nonumber\\
&& - \frac{r_+ - r_\ast}{r - r_+} \left( \frac{n_L + n_R}{4 \pi G} \partial_\phi - \frac{T_L + T_R}{4 G} \partial_t \right)^2,
\end{eqnarray}
where $G = n_L T_R - n_R T_L$.

By assuming $q_e = 0$ in~(\ref{scalarKNAdS}), one can find a hidden conformal symmetry from the radial equation. This leads to a holographic J-picture of a Kerr-Newman-AdS-dS black hole, as shown in~\cite{Chen:2010bh}. In this case, the radial equation could be written in terms of the $SL(2,R)$ quadratic Casimir as
\begin{equation}
\tilde{\cal H}^2 {\cal R}(r) = {\cal H}^2 {\cal R}(r) = -C {\cal R}(r).
\end{equation}
with the identification
\begin{eqnarray}
n^J_R = 0, & & n^J_L = - \frac{k}{2 (r_+ + r_\ast)},
\nonumber\\
T^J_R = \frac{k (r_+ - r_\ast)}{4 \pi a \Xi}, & & T^J_L = \frac{k (r^2_+ + r^2_\ast + 2 a^2)}{4 \pi a \Xi(r_+ + r_\ast)}. \label{identificationJ}
\end{eqnarray}
This helps us to fix the temperatures of dual CFT.

On the other hand, in the radial equation~(\ref{scalarKNAdS}), one may set $m = 0$ as well. This leads to another holographic description
of the Kerr-Newman-AdS-dS black hole, which will be called Q-picture. Define an operator $\partial_\chi$ acts on the ``internal space" of $U(1)$ symmetry of the complex scalar field and its eigenvalue is the charge of the scalar field $\partial_\chi \Phi = i \eta e \Phi$. Then the radial equation could be rewritten as
\begin{equation}
{\cal H}^2 {\cal R}(r) = - C {\cal R}(r),
\end{equation}
with the $SL(2,R)$ Casimir operator~(\ref{H2}) in which the derivative $\partial_\phi$ is replaced by $\partial_\chi$.
The temperatures in the dual 2D CFT, $T_{L, R}$, and $n_{L, R}$ and be identified as
\begin{eqnarray}
n^Q_L = - \frac{k (r_+ + r_\ast)}{4 (r_+ r_\ast - a^2)}, & & n^Q_R = - \frac{k (r_+ - r_\ast)}{4 (r_+ r_\ast - a^2)},
\nonumber\\
T^Q_L = \frac{k \eta(r^2_+ + r^2_\ast + 2 a^2)}{4 \pi q_e (r_+ r_\ast - a^2)}, & & T^Q_R = \frac{k \eta (r^2_+ - r^2_\ast)}{4 \pi q_e (r_+ r_\ast - a^2)}. \label{identificationQ}
\end{eqnarray}

It would be interesting to study the hidden conformal symmetry in the extremal limit. In this case, the radial equation~(\ref{scalarKNAdS}) reduces to
\begin{eqnarray} \label{scalarexKNAdS}
\partial_r (r - r_+)^2 \partial_r {\cal R}(r) + \frac{1}{k^2} \frac{2 [ \omega (r^2_+ + a^2) - m a \Xi - e q_e r_+ ] (2 \omega r_+ - e q_e)}{r - r_+} {\cal R}(r)
\nonumber\\
+ \frac{1}{k^2} \frac{[ \omega (r^2_+ + a^2) - m a \Xi - e q_e r_+ ]^2}{(r - r_+)^2} {\cal R}(r) + C {\cal R}(r) = 0.
\end{eqnarray}
For the extremal black hole, we should use the following conformal coordinates~\cite{Chen:2010fr}
\begin{eqnarray}
\omega^+ &=& \frac{1}{2} \left( \alpha_1 t + \beta_1 \phi - \frac{\gamma_1}{r - r_+} \right),
\nonumber\\
\omega^- &=& \frac{1}{2} \left( e^{2 \pi T_L \phi + 2 n_L t} - \frac{2}{\gamma_1} \right),
\label{confcor}\\
y &=& \sqrt{\frac{\gamma_1}{2 (r - r_+)}} e^{\pi T_L \phi + n_L t}. \nonumber
\end{eqnarray}
Then the corresponding $SL(2,R)$ quadratic Casimir is
\begin{equation}
{\cal H}^2 = \partial_r (\Delta \partial_r) - \left( \frac{\gamma_1 (2 \pi T_L \partial_t - 2 n_L \partial_\phi)}{\bar A (r - r_+)} \right)^2 - \frac{2 \gamma_1 (2 \pi T_L \partial_t - 2 n_L \partial_\phi)}{\bar A^2 (r - r_+)}(\beta_1 \partial_t - \alpha_1 \partial_\phi),
\end{equation}
where $\bar A = 2 \pi T_L \alpha_1 - 2 n_L \beta_1$ and $\Delta = (r - r_+)^2$.

In the J-picture, we set $q_e = 0$ and rewrite the radial equation as
\begin{equation}
\tilde{\cal H}^2 {\cal R}(r) = {\cal H}^2 {\cal R}(r) = -C {\cal R}(r).
\end{equation}
with the identification
\begin{equation}
\alpha^J_1 = 0, \qquad \beta^J_1 = \frac{\gamma_1 k}{a}, \qquad T^J_L = \frac{k}{\Xi} \frac{r_+^2 + a^2}{4 \pi a r_+}, \qquad n^J_L = - \frac{k}{4 r_+}.
\end{equation}
The left-temperature $T^J_L$ and $n^J_L$ are consistent with the identification~(\ref{identificationJ}) in the extremal limit. In the Q-picture, we set $m = 0$ and rewrite the radial equation as
\begin{equation}
{\cal H}^2 {\cal R}(r) = - C {\cal R}(r),
\end{equation}
with the identification
\begin{eqnarray}
& & \alpha^Q_1 = - \frac{k}{r_+^2 - a^2} \gamma_1, \qquad \beta^Q_1 = \frac{2 r_+ \eta k}{q_e (r^2_+ - a^2)} \gamma_1,
\nonumber\\
& & n^Q_L = - \frac{k r_+}{2 (r^2_+ - a^2)}, \qquad T^Q_L = \frac{\eta k (r^2_+ + a^2)}{2 \pi q_e (r^2_+ - a^2)}.
\end{eqnarray}
Once again, the left-temperature $T^Q_L$ and $n^Q_L$ are both consistent with the identification~(\ref{identificationQ}) in the extremal limit.

For a RN-AdS-dS black hole, corresponding to the limit $a = 0$, there is only one holographic description. The J-picture in such limit is actually singular.

In the limit of vanishing cosmological constant, we recover both the J-picture and Q-picture of the asymptotically flat
case, studied in~\cite{Chen:2010yw}.

\section{Microscopic description in Q-picture}

In order to have a complete Q-picture description, we need to determine the central charges of the dual CFT. Let us consider the near horizon geometry of the extreme Kerr-Newman-AdS-dS black hole with degenerated horizon at $r_+ = r_* = r_0$. In terms of the following new coordinates
\begin{equation} \label{coordinate}
r = r_0 + \epsilon \hat r, \qquad t = \frac{r_0^2 + a^2}{k} \frac{\hat t}{\epsilon}, \qquad \phi = \hat \phi + \Omega_H t,
\end{equation}
the metric becomes~\cite{Hartman:2008pb, Rasmussen:2010xd}
\begin{equation} \label{NHmetric}
ds^2 = \Gamma(\theta) \left[ - \hat r^2 d\hat t^2 + \frac{d\hat r^2}{\hat r^2} + \alpha(\theta) d\theta^2 \right] + \gamma(\theta) (d\hat \phi + p^J \, \hat r d\hat t)^2,
\end{equation}
where
\begin{equation}
\Gamma(\theta) = \frac{\rho_0^2}{k}, \qquad \alpha(\theta) = \frac{k}{\Delta_\theta}, \qquad \gamma(\theta) = \frac{(r_0^2 + a^2)^2 \Delta_\theta \sin^2\theta}{\Xi^2 \, \rho_0^2},
\end{equation}
and
\begin{equation}
\rho_0^2 = r_0^2 + a^2 \cos^2\theta, \qquad p^J = \frac{2 a r_0 \Xi}{k (r_0^2 + a^2)}.
\end{equation}
The near horizon gauge field is
\begin{equation}
A_{[1]} = \frac{q_e}{\rho_0^2} \left( \frac{r_0 a \sin^2\theta}{\Xi} d\hat\phi + \frac{r_0^2 - a^2 \cos^2\theta}{k} \hat r d\hat t \right) + f(\theta) (d\hat \phi +  p^J \, \hat r d\hat t),
\end{equation}
with
\begin{equation}
f(\theta) = \frac{q_m (r_0^2 + a^2)}{\Xi \rho_0^2} \cos\theta.
\end{equation}
The central charges of the dual CFT could be read from the near-horizon geometry of the extremal black holes. It turns out that in the J-picture ($q_e = q_m = 0$, i.e. $q = 0$), the dual 2D CFT has the central charges
\begin{equation}
c = 3 p^J \int_0^{\pi} d\theta \sqrt{\Gamma(\theta) \alpha(\theta) \gamma(\theta)} = \frac{12 a r_0}{k}.
\end{equation}
For a general expression of the central charge, one should rewrite $r_0$ in the notion of mass, i.e. $(r_+ + r_\ast)/2$, as it has done in the Kerr black holes. Consequently, the J-picture left- and right-handed central charges are
\begin{equation}
c^J_L = c^J_R = \frac{6 a (r_+ + r_\ast)}{k}.
\end{equation}
It is easy to see from the Cardy formula
\begin{equation}
S^J_{CFT} = \frac{\pi^2}{3} (c^J_L T^J_L + c^J_R T^J_R),
\end{equation}
that the CFT entropy recovers exactly the macroscopic Bekenstein-Hawking entropy. This provides a primary evidence to our holographic picture.

In order to get the central charges in Q-picture, ($a = 0$), we need to uplift the geometry to 5D. Combine the $U(1)$ gauge bundle
\begin{equation}
A_{[1]} = p^Q \hat r d\hat t + \frac{q_m}{\Xi} \cos\theta d\hat \phi, \qquad p^Q = \frac{q_e}{k},
\end{equation}
with the geometry and write the 5D space as
\begin{equation}
ds^2 = ds^2_{BH} + (d y + A_{[1]})^2,
\end{equation}
where $y$ is the fiber coordinate with period $2 \pi \eta$ and $ds^2_{BH}$ is the 4D near horizon metric~(\ref{NHmetric}). We can choose similar boundary conditions as in~\cite{Hartman:2008pb}
\begin{equation}
h_{\mu\nu} \sim \left(\begin{array}{ccccc}
  \hat r^2 & \hat r & 1/\hat r & 1/\hat r^2 & 1 \\
  & 1/\hat r & 1 & 1/\hat r & 1 \\
  & & 1/\hat r & 1/\hat r^2 & 1/\hat r \\
  & & & 1/\hat r^3 & 1/\hat r \\
  & & & & 1
  \end{array} \right),
\end{equation}
in the basis of $(\hat t, \hat \phi, \theta, \hat r, y)$. The most general diffeomorphisms preserving those boundary conditions are
\begin{equation}
\zeta^{(y)} = \epsilon(y) \partial_{y} - \hat r \epsilon^\prime(y) \partial_{\hat r},
\end{equation}
where $\epsilon(y) = e^{-i n y}$. The central charge can be computed from the 5D generalization of the treatment in~\cite{Hartman:2008pb}. It turns out that the central charges associated with $\zeta^{(y)}$ is
\begin{equation}\label{centralQ}
c = \frac{3 p^Q}{\eta} \int_0^{\pi} d\theta \sqrt{\Gamma(\theta) \alpha(\theta) \gamma(\theta)} = \frac{6 q_e r_0^2}{\eta \Xi k}.
\end{equation}
Similarly, one should rewrite $r_0$ in terms of general variables, $r_+, r_\ast$ and $a$, to have general central charges. For the Q-picture, inspired by the results of the RN black hole, one should rewrite $r_0^2$ in the notion of charge square, i.e. $r_+ r_\ast - a^2$, which leads to the general expression of Q-picture left- and right-handed central charges
\begin{equation}
c^Q_L = c^Q_R = \frac{6 q_e}{\eta \Xi} \frac{r_+ r_\ast - a^2}{k}.
\end{equation}

In the extreme limit $r_+ = r_\ast$, we have
\begin{equation}
a^2 = \frac{r_+^2 (1 + 3 r_+^2/l^2) - q^2}{1 - r_+^2/l^2}\,.
\end{equation}
Taking $a \to 0$, we see that the central charge~(\ref{centralQ}) recover precisely the one of the extreme Reissner-Nordstr\"om-AdS-dS black holes~\cite{Hartman:2008pb},
\begin{equation}
c \to \frac{6 q_e \tilde r_0^2}{\eta}\,,
\end{equation}
where
\begin{equation} \label{ceRNAdS}
\tilde r_0^2 \to \frac{r_+^2 (1 - r_+^2/l^2)}{1 + 6 r_+^2/l^2 - 3 r_+^4/l^4 - q^2/l^2}\,.
\end{equation}
We conclude that in the Q-picture, the Kerr-Newman-AdS-dS black hole is described by a dual 2D CFT with the central charges~(\ref{centralQ}) and temperatures~(\ref{identificationQ}).

\subsection{Thermodynamics}

One can check that in the Q-picture the Bekenstein-Hawking entropy of a Kerr-Newman-AdS-dS black hole could be reproduced through the Cardy formula using~(\ref{identificationQ}) and~(\ref{centralQ})
\begin{equation}
S_{BH} = \frac{\pi (r_+^2 + a^2)}{\Xi} \equiv \frac{\pi^2}{3} (c_L T_L + c_R T_R) = S_{CFT}.
\end{equation}
This provides a nontrivial check for our suggestion.

From the first law of thermodynamics
\begin{equation}
\delta S_{BH} = \frac{\delta M - \Omega_H \delta J - \Phi_e \delta q}{T_H} = \frac{\delta E_L}{T_L} + \frac{\delta E_R}{T_R},
\end{equation}
we always have
\begin{equation}
\delta E_L = \omega_L - q_L \mu_L, \qquad \delta E_R = \omega_R - q_R \mu_R,
\end{equation}
in both pictures. But the identifications are quite different. The derivations correspond to the parameters of probe scalar field by $\delta M = \omega, \; \delta J = m, \; \delta Q = e$. In the J-picture, we have
\begin{eqnarray} \label{identificationJ1}
&& \omega^J_L = \frac{r^2_+ + r_\ast^2 + 2 a^2}{2 a \Xi} \omega, \qquad \omega^J_R = \frac{r^2_+ + r_\ast^2 + 2 a^2}{2 a \Xi} \omega - m,
\nonumber\\
&& q^J_L = q^J_R = \delta Q = e,
\nonumber\\
&& \mu^J_L = \frac{q (r^2_+ + r_\ast^2 + 2 a^2)}{2 a \Xi (r_+ + r_\ast)}, \qquad \mu^J_R = \frac{q (r_+ + r_\ast)}{2 a \Xi}.
\end{eqnarray}
While in the Q-picture we have
\begin{eqnarray} \label{identificationQ1}
&& \omega^Q_L = \frac{\eta (r_+ + r_\ast) (r_+^2 + r_\ast^2 + 2a^2)}{2 q_e (r_+ r_\ast - a^2)} \omega,
\nonumber\\
&& \omega^Q_R = \frac{\eta (r_+ + r_\ast) (r_+^2 + r_\ast^2 + 2a^2)}{2 q_e (r_+ r_\ast - a^2)} \omega - \frac{\eta a \Xi (r_+ + r_\ast)}{q_e (r_+ r_\ast - a^2)} m,
\nonumber\\
&& q^Q_L = q^Q_R = \delta Q = e,
\nonumber\\
&& \mu^Q_L = \frac{\eta (r_+^2 + r_\ast^2 + 2 a^2)}{2 (r_+ r_\ast - a^2)}, \qquad \mu^Q_R = \frac{\eta (r_+ + r_\ast)^2}{2 (r_+ r_\ast - a^2)}.
\end{eqnarray}
In the limit of vanishing cosmological constant, these quantities reduce to the ones found in~\cite{Chen:2010yw}.

\subsection{Superradiant scattering}

In a 2D conformal field theory, one can define the two-point function
\begin{equation}
G(t^+, t^-) = \langle {\cal O}^\dagger_\phi(t^+, t^-) {\cal O}_\phi(0) \rangle,
\end{equation}
where $t^+, t^-$ are the left and right moving coordinates of 2D worldsheet and ${\cal O}_\phi$ is the operator corresponding to the field perturbing the black hole. For an operator of dimensions $(h_L, h_R)$, charges $(q_L, q_R)$ at temperatures $(T_L, T_R)$ and chemical potentials $(\mu_L, \mu_R)$, the two-point function  is dictated by conformal invariance and takes the form~\cite{Cardy:1984bb}:
\begin{equation} \label{G-Mink}
G(t^+, t^-) \sim (-1)^{h_L + h_R} \left( \frac{\pi T_L}{\sinh(\pi T_L t^+)} \right)^{2h_L} \left( \frac{\pi T_R}{\sinh(\pi T_R t^-)} \right)^{2h_R} e^{i q_L \mu_L t^+ + i q_R \mu_R t^-}.
\end{equation}
The retarded correlator $G_R (\omega_L, \omega_R)$ is analytic on the upper half complex $\omega_{L,R}$-plane and its value along the positive imaginary $\omega_{L,R}$-axis gives the Euclidean correlator:
\begin{equation} \label{GER}
G_E(\omega_{L, E}, \omega_{R, E}) = G_R(i \omega_{L,E}, i \omega_{R,E}), \qquad \omega_{L, E}, \; \omega_{R, E} > 0.
\end{equation}
At finite temperature, $\omega_{L,E}$ and $\omega_{R,E}$ take discrete values of the Matsubara frequencies
\begin{equation}
\omega_{L, E} =  2 \pi m_L T_L, \qquad \omega_{R,E} = 2 \pi m_R T_R,
\end{equation}
where $m_L, m_R$ are integers for bosonic modes and are half integers for fermionic modes.

In a 2D CFT, the Euclidean correlator $G_E$ is obtained by a Wick rotation $t^+ \to i \tau_L$, $t^- \to i \tau_R$, and is determined by the conformal symmetry. At finite temperature the Euclidean time is taken to have period $2 \pi/T_L, 2 \pi/T_R$ and via analytic continuation the momentum space Euclidean correlator is given by~\cite{Maldacena:1997ih}
\begin{eqnarray} \label{GE}
G_E(\omega_{L, E}, \omega_{R, E}) &\sim& T_L^{2 h_L - 1} T_R^{2 h_R - 1} \, e^{i \frac{\bar\omega_{L, E}}{2 T_L}} \, e^{i \frac{\bar\omega_{R, E}}{2 T_R}} \; \Gamma\left( h_L + \frac{\bar\omega_{L, E}}{2 \pi T_L} \right) \Gamma\left( h_L - \frac{\bar\omega_{L, E}}{2 \pi T_L} \right)
\nonumber\\
&& \times \Gamma\left( h_R + \frac{\bar\omega_{R, E}}{2 \pi T_R} \right) \Gamma\left( h_R - \frac{\bar\omega_{R, E}}{2 \pi T_R} \right),
\end{eqnarray}
where
\begin{equation}
\bar\omega_{L, E} = \omega_{L,E} - i q_L \mu_L, \qquad \bar\omega_{R, E} = \omega_{R, E} - i q_R \mu_R.
\end{equation}

Since the function $\Delta_r$ is quartic, the radial equation for generic black hole is intractable. In the study of the hidden conformal symmetry, we focus on the near horizon region. However, when we try to discuss the scattering issue and move away from the horizon, the expansion of $\Delta_r$ to the second order of $(r - r_+)$ breaks down. Therefore in general, we cannot use the radial equation~(\ref{scalarKNAdS}) to discuss the scattering process, even though it provides useful information on dual CFTs. Nevertheless, for near-extremal Kerr-Newman-AdS-dS black hole, we may pose a
well-defined scattering problem. To this end, we need to zoom in the near-horizon region and introduce the coordinates~(\ref{coordinate}) to describe the geometry. In this case, we have to focus on the frequencies near the superradiant bound $\omega_s$,
\begin{equation}
\omega = \omega_s + \hat \omega \frac{\epsilon}{r_0}\,, \qquad \omega_s = m \Omega_H + e \Phi_e,
\end{equation}
where $\Omega_H$ and $\Phi_e$ are the horizon angular velocity and the electric potential given in~(\ref{OH}) and~(\ref{Epotential}) respectively.

The wave function of the radial equation is then
\begin{equation}
{\cal R}(z) = z^{\alpha} (1 - z)^{\beta} F(a, b, c\,; z),
\end{equation}
with \, $z = \frac{\hat r - \lambda}{\hat r + \lambda}$ \, and
\begin{eqnarray}
&& \alpha = -i \hat A, \qquad \beta = \frac{1}{2} \left( 1 - \sqrt{1 - 4 \hat C} \right),
\\
&& c = 1 + 2 \alpha, \qquad a = \alpha + \beta + i \hat B, \qquad b = \alpha + \beta - i \hat B,
\\
&& \hat A = \frac{\hat \omega}{2 \lambda}, \quad \hat B = \frac{\hat \omega}{2 \lambda} - \frac{2 m \Omega_H r_+}{k} - \frac{e q}{k} \frac{r^2_+ - a^2}{r^2_+ + a^2}, \quad \hat C = \hat C(\omega_s).
\end{eqnarray}
The solution behaves asymptotically as
\begin{equation}
{\cal R}(r) \simeq A_1 r^{h-1} + A_2 r^{-h},
\end{equation}
where $h$ is the conformal weight of the scalar field
\begin{equation}
h = 1 - \beta = \frac{1}{2} \left( 1 + \sqrt{1 - 4 \hat C} \right).
\end{equation}
Taken $A_1$ as the source and $A_2$ as the response, the retarded Green's function is just~\cite{Chen:2010xu}
\begin{eqnarray}
G_R &\sim& \frac{A_2}{A_1}
\nonumber\\
&=& \frac{\Gamma(1 - 2h)}{\Gamma(2h - 1)} \frac{\Gamma\left( h - i (\hat A - \hat B) \right) \Gamma\left( h - i (\hat A + \hat B) \right)}{\Gamma\left( 1 - h - i (\hat A - \hat B) \right) \Gamma\left( 1 - h - i (\hat A + \hat B) \right)}
\nonumber\\
&=& \frac{\Gamma(1 - 2h)}{\Gamma(2h - 1)} \frac{\Gamma\left( h - i \frac{\omega_L - q_L \mu_L}{2 \pi T_L} \right) \Gamma\left( h - i \frac{\omega_R - q_R \mu_R}{2 \pi T_R} \right)}{\Gamma\left( 1 - h - i \frac{\omega_L - q_L \mu_L}{2 \pi T_L} \right) \Gamma\left( 1 - h - i \frac{\omega_R - q_R \mu_R}{2 \pi T_R} \right)}. \label{realtime}
\end{eqnarray}
In the last line, we have applied the identification~(\ref{identificationJ}) and~(\ref{identificationJ1}) in J-picture and the identification~(\ref{identificationQ}) and~(\ref{identificationQ1}) in Q-picture. We see that in both pictures, the real-time correlator~(\ref{realtime}) is in perfect match with the CFT prediction. So is the absorption cross section.

Note that in two different pictures, though the retarded Green's functions share the same expression~(\ref{realtime}), the temperatures, chemical potentials and frequencies are different. However the conformal weights are the same, reflecting the fact the conformal weight is closely related to the asymptotical behavior.

\section{Discussions}

In this paper, we showed that there are two different holographic descriptions of a generic non-extremal Kerr-Newman-AdS-dS black hole. One is called J-picture, whose construction is based on the black hole angular momentum. The other one is called Q-picture, whose construction originate from the electric charge of the black hole. In these two different pictures, neither the central charges nor the temperatures of the dual CFTs is the same. In particular, in the Q-picture, the central charges and the temperatures are parameterized by a varying constant $\eta$, sharing the same feature in the holographic descriptions of RN and Kerr-RN black holes. As a byproduct of our analysis, we showed that there exists only one holographic description for a generic nonextremal RN-AdS-dS black hole.

The discussion in this paper could be generalized easily in several directions. It is straightforward to consider a dyonic Kerr-Newman-AdS-dS black hole. Also the analysis could be applied to the higher dimensional Kerr-AdS-dS black holes and multi-charged black holes, both of which are expect to have multiple holographic descriptions. Moreover, the existence of multiple dual pictures leads to a natural believe that there should exist a certain duality among those different pictures. It is desirable to clarify this important issue.

Very recently it was pointed out in \cite{Guica:2010ej} that the existence of different holographic dual descriptions of 5D Kerr-Newman black hole is related to the long and short string pictures. It would be interesting to see if there exist similar pictures in our case.

\section*{Acknowledgements}
CMC is grateful to Institute of Theoretical Physics and Morningside Center of Mathematics, Chinese Academy of Sciences for the hospitality when this paper was initiated.
The work of BC was partially supported by NSFC Grant No.10775002, 10975005.
This work by CMC was supported by the National Science Council of the R.O.C. under the grant NSC 99-2112-M-008-005-MY3 and in part by the National Center of Theoretical Sciences (NCTS).

\end{document}